\documentclass[aps,prl,superscriptaddress,11pt]{revtex4-1}
\usepackage{graphicx}
\usepackage{amsmath}
\usepackage{color}
\usepackage{soul}

\begin{document}
\title{Exploring the phase diagram of the two-impurity Kondo problem}
\author{A. Spinelli}
\author{M. Gerrits}
\author{R. Toskovic}
\author{B. Bryant}
\affiliation{Department of Quantum Nanoscience, Kavli Institute of Nanoscience, Delft University of Technology, Lorentzweg 1, 2628 CJ Delft, The Netherlands}
\author{M. Ternes}
\affiliation{Max-Planck Institute for Solid State Research, Heisenbergstra{\ss}e 1, 70569 Stuttgart, Germany}
\author{A. F. Otte}\thanks{a.f.otte@tudelft.nl}
\affiliation{Department of Quantum Nanoscience, Kavli Institute of Nanoscience, Delft University of Technology, Lorentzweg 1, 2628 CJ Delft, The Netherlands}

\begin{abstract}
A system of two exchange-coupled Kondo impurities in a magnetic field gives rise to a rich phase space hosting a multitude of correlated phenomena. Magnetic atoms on surfaces probed through scanning tunnelling microscopy provide an excellent platform to investigate coupled impurities, but typical high Kondo temperatures prevent field-dependent studies from being performed, rendering large parts of the phase space inaccessible. We present an integral study of pairs of Co atoms on insulating Cu$_2$N/Cu(100), which each have a Kondo temperature of only 2.6~K. In order to cover the different regions of the phase space, the pairs are designed to have interaction strengths similar to the Kondo temperature. By applying a sufficiently strong magnetic field, we are able to access a new phase in which the two coupled impurities are simultaneously screened. Comparison of differential conductance spectra taken on the atoms to simulated curves, calculated using a third order transport model, allows us to independently determine the degree of Kondo screening in each phase.
\end{abstract}
\maketitle

The coupling of individual magnetic atoms to the itinerant host electrons of a metal substrate can lead to the formation of a correlated Kondo state in which the magnetic moment is effectively reduced \cite{hewson1997kondo}. This has been shown for 3$d$-atoms on bare metal surfaces \cite{Madhavan1998,Li1998} as well as on thin decoupling layers \cite{Otte2008,vonBergmann2015} and leads to a strong spectroscopic feature at the Fermi energy. Pairs of magnetic atoms are considerably more complex because in addition to the Kondo coupling they can also couple to each other through exchange interactions mediated by the substrate electrons. Depending on their spatial distance, this oscillatory Rudermann-Kittel-Kasuya-Yoshida (RKKY) interaction results in a ferromagnetic (FM) or antiferromagnetic (AFM) coupling \cite{Ruderman1954, Kasuya1956, Yosida1957}. 

\begin{figure}[htbp]
\begin{centering}
\includegraphics{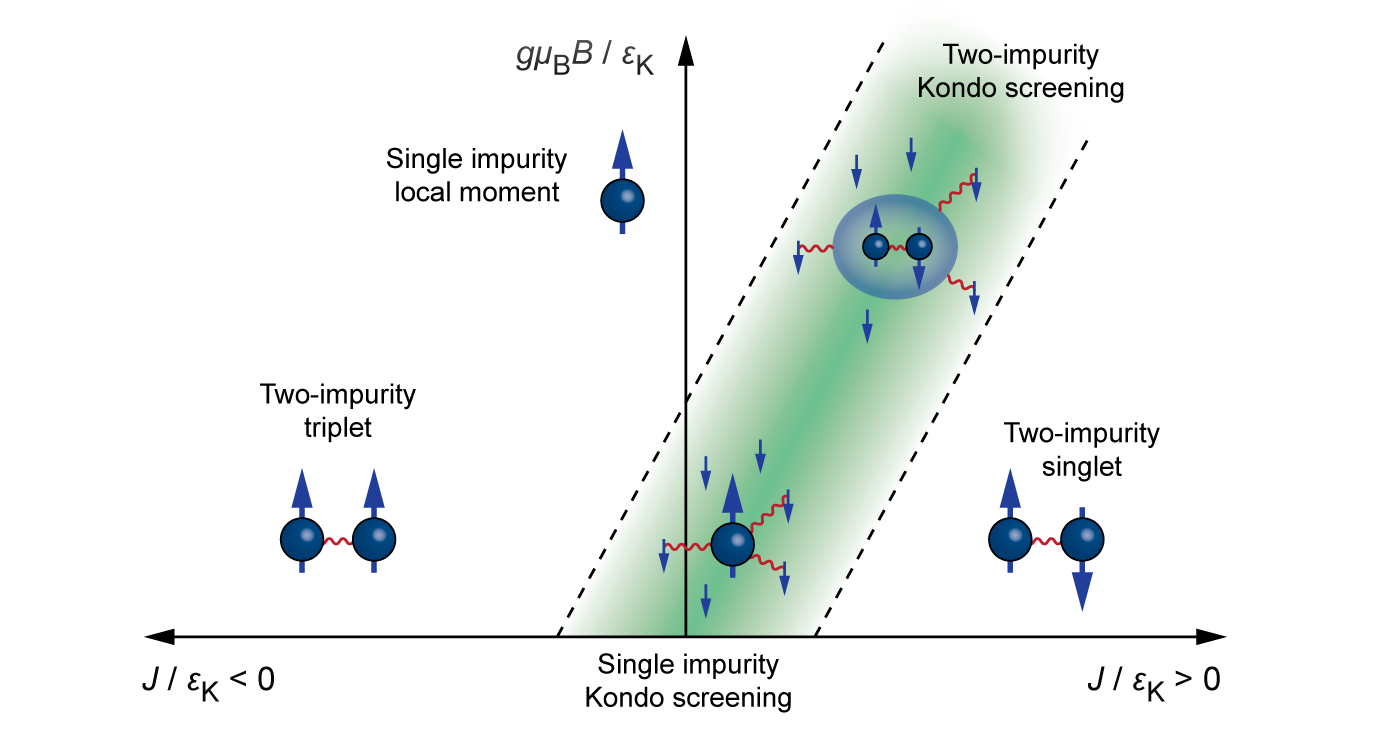}
\caption{\textbf{Phase diagram of the two-impurity Kondo problem.} Schematic phase diagram of two coupled Kondo-screened spins with varying interaction strength $J$ and external field $B$ transverse to the main anisotropy axis of the individual spins. When $|J|$ is small compared to the characteristic Kondo energy $\epsilon_{\rm K}$, at $B=0$ the two spins are independently screened by the substrate electrons, while for $|J|\gg\epsilon_{\rm K}$ a non-magnetic singlet or high-spin triplet state forms. For $J>0$ a sufficient $B$-field can lead to the formation of a new, combined correlated state in which both spins are screened.}
\label{Figure1}
\end{centering}
\end{figure}

As shown in Fig.~\ref{Figure1}, the competition between these two effects in combination with an external magnetic field embodies rich physics ranging from a correlated singlet or triplet state to complex Kondo states and has been of considerable theoretical interest for decades \cite{Jayaprakash1981, Jones1987, Jones1988, Jones1989, Silva1996, Simon2005, Zitko2010, Jabben2012}. The lower part of this phase space has previously been investigated experimentally through studies at zero magnetic field on coupled quantum dots \cite{Craig2004}, molecules \cite{Wegner2009, Tsukahara2011}, and atoms on top of \cite{Jamneala2001, Wahl2007, Neel2011} as well as buried below \cite{Pruser2014} a metal surface. In particular, controlled mechanical separation of two Kondo atoms enabled continuous tuning between the single impurity Kondo screening and two-impurity singlet phases \cite{Bork2011}. However, field-dependent measurements have so far been hampered by high Kondo temperatures for atoms in direct contact with a metal, resulting in wide resonances that are impossible to split or recombine with experimentally available magnetic fields. Kondo temperatures can be reduced by decoupling the atoms from their host either by means of molecular ligands \cite{Tsukahara2011, Dubout2015} or a thin insulating layer \cite{Otte2008}.

Here we use a low-temperature scanning tunnelling microscope (STM) in ultra-high vacuum to assemble pairs of Kondo-screened Co atoms on a thin insulating Cu$_2$N/Cu(100) substrate. By adjusting the relative position and orientation of the atoms on the underlying crystal lattice we are able to tune their exchange interaction strength, and by applying magnetic fields up to 8~T we can access the complete phase space of Fig.~\ref{Figure1}. Furthermore, by simulating the spectra using a transport model accounting only for weak Kondo coupling, we independently estimate the Kondo screening. We find that when in AFM coupled dimers the field exactly cancels the exchange interaction, Kondo resonances are re-established, signifying a phase transition through the two-impurity Kondo-screened phase.

\section{Results}
A single Co atom on Cu$_2$N can be described by an effective spin $S=3/2$, and the crystal field induced by the surface is such that the two lowest energy states have magnetization $m=\pm1/2$, degenerate in absence of external magnetic field \cite{Otte2008}. Its differential conductance (${\rm d}I/{\rm d}V$) spectrum shows a sharp peak at zero excitation voltage, corresponding to a Kondo resonance with a Kondo temperature $T_{\rm K}=2.6 \pm 0.2$~K \cite{Otte2008}, equivalent to a characteristic Kondo energy $\varepsilon_{\rm K}=k_{\rm B}T_{\rm K}=0.22\pm0.02$~meV, which is much lower than for single atoms directly on  \cite{Madhavan1998,Li1998} or inside a metal \cite{Pruser2014}. This peak splits when a magnetic field is applied due to the Zeeman effect removing the degeneracy of the states with $m=\pm1/2$ \cite{vonBergmann2015}. If a second magnetic atom is placed adjacent to the first, the Kondo peak splits similarly to when an effective magnetic field is applied; the intensity of this field scales with the strength of the exchange interaction between the two atoms \cite{Otte2009}. 

To model the differential conductance spectra we use a single spin anisotropy Hamiltonian in presence of an external magnetic field with energy $\mu_{\rm B}{\bf B}$ ($\mu_{\rm B}$ being the Bohr magneton), and we assume isotropic Heisenberg exchange $J$ \cite{Fernandez-Rossier2009} between the atoms in the dimer:
\begin{equation}\label{H}
\mathcal{H}=J{\bf S}^{\rm A}\cdot{\bf S}^{\rm B}-\sum_{\substack{i,\mu}}\left(\lambda^2\Lambda_{\mu\mu}S_{\mu}^{i}S_{\mu}^{i}+2(1-\lambda\Lambda_{\mu\mu})\mu_{\rm B} B_{\mu}S_{\mu}^{i}\right).  
\end{equation} 
Here the parameters $\Lambda_{\mu\mu}$ ($\mu=x,y,z$) come from second order expansion of the spin-orbit coupling $\lambda {\bf L}^i\cdot {\bf S}^i$ \cite{Dai2008} (with spin-orbit constant $\lambda=-21$~meV) and represent the extent to which the angular momentum $\mathbf{L}$ (which for a free Co atom has magnitude $L=3$) is unquenched along the $\mu$-direction, while the indices $i={\rm A},{\rm B}$ refer to the different atoms in the dimer. Individual Co atoms experience a strong hard-axis anisotropy $\Lambda_{xx}=0$, $\Lambda_{yy}=\Lambda_{zz}=6.4~{\rm eV}^{-1}$ along the vacancy direction ($x$) of the Cu$_2$N lattice (shown in Fig.~\ref{Figure2}e), as a result of which the $m=\pm3/2$ doublet is split off by $2\lambda^2 \Lambda_{zz}=5.6$~meV from the lower energy $m=\pm1/2$ doublet \cite{Otte2008}. When written in terms of the conventional phenomenological anisotropy parameters $D$ and $E$ \cite{Hirjibehedin2007,Brune2009,Gatteschi2006}, this corresponds to $D=+2.8$~meV, $E=0$. In most dimers studied here, the intra-dimer exchange coupling $J$, which can be either antiferromagnetic (AFM, $J>0$) or ferromagnetic (FM, $J<0$), is much weaker than the anisotropy energy ($|J|\ll\lambda^2\Lambda_{zz}$). As a result, each dimer is characterised by a low energy quartet that consists of combinations of $m=\pm 1/2$ states of both atoms.

\subsection{Exchange interaction}
\begin{figure}[t!]
\begin{centering}
\includegraphics{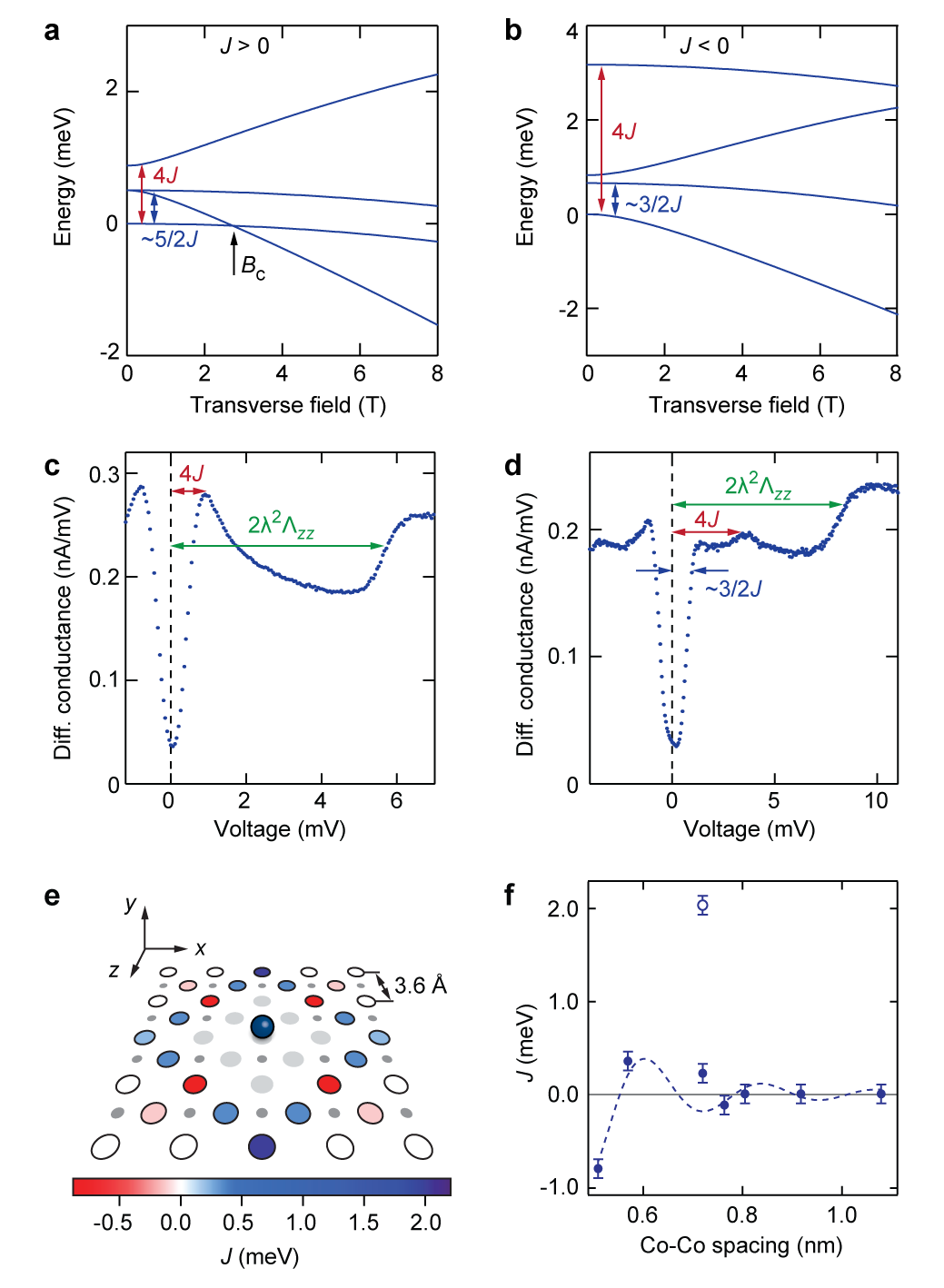}
\caption{\textbf{Sign and strength of the exchange interaction between atoms.} (\textbf{a--b}) Energy vs transverse magnetic field of the four lowest energy states for two example dimers, corresponding to the instances presented in panels \textbf{c--d}, with AFM ($J>0$) and FM ($J<0$) coupling respectively. In the AFM case, a state crossing is expected at a field $B_{\rm c}$. (\textbf{c--d}) Example of zero-field differential conductance spectra taken on a Co atom in two different dimers. Red and green arrows show respectively the exchange and anisotropy energies. In the FM spectrum, an additional step can be resolved at $3/2J$. (\textbf{e}) Colour map of the experimentally extracted coupling strength between Co atoms on Cu$_2$N. Each circle corresponds to the position of the second Co atom atop a Cu atom of the lattice, with respect to a reference Co atom (blue sphere). Light gray represent Cu positions too close to allow a Co dimer to be built while the small grey circles are the N atoms. The axes identify the coordinate system for the reference atom, with the nitrogen direction defined as $z$-axis. (\textbf{f}) Coupling strength vs. inter-dimer distance. For comparative purposes, an isotropic three-dimensional RKKY curve with Fermi wavelength of bulk Cu and horizontal offset of $0.15$~nm, corresponding to a phase shift of 1.3$\pi$, is shown (dashed line). The point represented with an open circle is the $\left\{2,0\right\}$ dimer, whose coupling cannot be explained in terms of RKKY interaction only.}
\label{Figure2}
\end{centering}
\end{figure}

For either sign of $J$, the zero-field energy difference between the lowest and the highest state of the quartet is $4J$ (Figs.~\ref{Figure2}a--b). In the limit of $J$ being much smaller than the magnetic anisotropy, the energy difference between the ground state and the first excited state is $\sim(5/2)J$ or $\sim(3/2)J$ for AFM or FM coupling respectively. Figs.~\ref{Figure2}c--d show two ${\rm d}I/{\rm d}V$ spectra taken on Co atoms in two different dimers, where each step-like increase in the differential conductance corresponds to a spin excitation from the ground state to an excited state, and in which the two energy scales $4J$ and $2\lambda^2\Lambda_{zz}$ are clearly distinguishable. The evolution of the energetically lowest excitation as a function of magnetic field (in our experiment applied along the transverse direction $y$) reveals the sign of $J$. For AFM coupling, the step-energy decreases until the critical field $B_{\rm c} =(13/8)\times J/(g\mu_{\rm B})$ is reached, at which the ground state and the first excited state become degenerate (Fig.~\ref{Figure2}a). At $B>B_{\rm c}$ these states change their order and their energy difference increases with field. For FM coupling, the step-energy will only increase with increasing field and a crossing between the ground state and the first excited state does nor occur (Fig.~\ref{Figure2}b).

Following these considerations, we can extract the coupling value between any two atoms placed on the Cu$_2$N surface through differential conductance spectroscopy. In Fig.~\ref{Figure2}e we show the obtained coupling map with respect to a reference Co atom drawn in the center. Depending on the relative distance and orientation between the atoms in the pair, that can be placed only in specific positions allowed by the substrate, the coupling can be tuned in strength and sign. The obtained results are in good agreement with the coupling values that were previously extracted for Fe dimers \cite{Bryant2013}. 

In previous experiments \cite{Khajetoorians2012} and theoretical calculations \cite{Simon2005,Simon2011} on magnetic atoms directly on a metal surface, the dominant exchange mechanism was found to be RKKY interaction \cite{Ruderman1954,Kasuya1956,Yosida1957}. It is not known to what extent this is still true in the current case, where the adatoms are embedded in a covalently bonded network \cite{Hirjibehedin2007} which decouples them from the substrate conduction electrons mediating this interaction. In Fig.~\ref{Figure2}f we plot the extracted coupling value of each dimer as a function of the absolute separation distance. We classify each dimer by the number of unit cells separating the Co atoms in the two symmetry directions of the underlying Cu$_2$N lattice, first stating the distance along the nitrogen direction ($z$) and then along the vacancy direction ($x$): $\{\rm N,\rm v\}$. For all dimers studied except the $\left\{2,0\right\}$ type, depicted with a different symbol, the coupling dependence on the distance seems to be qualitatively comparable to isotropic bulk RKKY \cite{kittel1963quantum}, shown in the same figure for comparative purposes only. A full theoretical treatment of the RKKY coupling on this composite surface, in which also anisotropic RKKY interaction \cite{Zhou2010} could play a role, is beyond the scope of this paper.

In the case of the $\left\{2,0\right\}$ dimer, the exchange coupling is an order of magnitude larger than for the equally spaced $\left\{0,2\right\}$ dimer. According to isotropic RKKY theory the coupling strength and sign depend only on absolute distance, and should therefore be identical for these two dimers. In this particular case we believe that, in addition to RKKY interaction, superexchange coupling plays an important role \cite{Koch2012}. Being mediated by the Co--N and Cu--N bonds separating the two Co atoms, superexchange coupling should be strongly direction-dependent: every $90^{\circ}$~corner in the coupling path significantly reduces its magnitude \cite{Goodenough1958,Kanamori1959}. Of all dimers studied, the $\left\{2,0\right\}$ geometry is the only case where the two Co atoms are connected by a series of bonds without any corner. To further investigate the role of superexchange interaction, dimers would have to be built in which Co atoms are separated by a single N atom only, which so far we have not been able to do controllably.

\subsection{Spectroscopy measurements}
\begin{figure}[htbp]
\begin{centering}
\includegraphics[width=\textwidth]{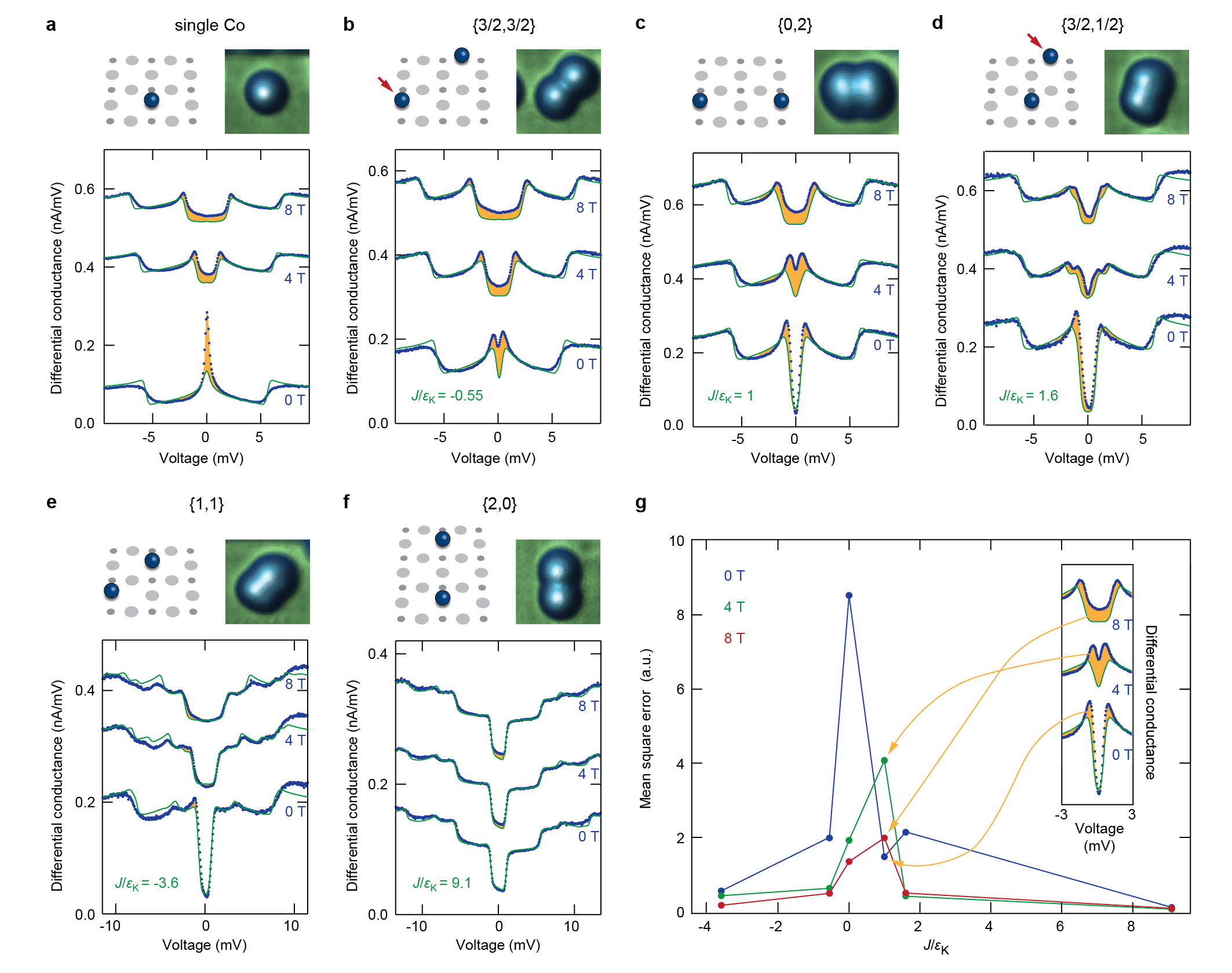}
\caption{\textbf{Overview of dimer configurations: experimental data and simulated spectra.} (\textbf{a--f}) Measured ${\rm d}I/{\rm d}V$ spectra (blue dots) and simulated curves (green lines) on a single Co atom and on all five types of dimers showing a measurable coupling, presented with increasing value of $|J/\varepsilon_{\rm K}|$, with $\varepsilon_{\rm K}=0.22\pm 0.02$~meV, for zero magnetic field and fields applied in the $y$ direction, perpendicular to the sample surface. Field spectra have been offset for clarity; all measurements were performed at 330~mK. Above each panel, a corresponding atomic lattice diagram and topographic STM image of the measured dimer are shown. Large and small grey circles represent respectively Cu and N atoms; the blue spheres the Co atoms. The arrows in panels b and d indicate which atom of the dimer is presented in the figure; those are the only two cases in which the two atoms are not equivalent. (\textbf{g}) Mean square error between the data and the simulated conductance curve as a function of $J/\varepsilon_{\rm K}$ for zero field (blue), 4~T (green) and 8~T (red).}
\label{Figure3}
\end{centering}
\end{figure}

We measured ${\rm d}I/{\rm d}V$ spectra on every atom of each dimer at different values of external magnetic field applied perpendicular to the sample plane, transverse to the magnetic hard axis (Figs.~\ref{Figure3}b--f). For comparison, spectra taken on a single Co are shown in Fig.~\ref{Figure3}a. Additionally, we simulate the experimental data by diagonalizing the Hamiltonian (\ref{H}) and using a perturbative scattering model in which the tunnelling electrons interact with the localized spin $\mathbf{S}$ via an exchange interaction $\mathbf{S}\cdot$\boldmath${\sigma}$, with $\mathbf{\sigma}$\unboldmath$=(\sigma_x,\sigma_y,\sigma_z)$ the standard Pauli matrices for spin-1/2 electrons \cite{LothNature2010}. In the differential conductance curves of Figs.~\ref{Figure3}b--e, the magnetic anisotropy is best described with $\Lambda_{xx}=0$, and $\Lambda_{yy}$ and $\Lambda_{zz}$ varying between 6.1~eV$^{-1}$ and 7.0~eV$^{-1}$, very close to the single Co values. However, this is not the case for the $\left\{2,0\right\}$ dimer, presented in Fig.~\ref{Figure3}f, for which $\Lambda_{zz}=2\Lambda_{yy}$. We believe that in this structure the strain induced by the presence of the second atom is playing a critical role, enhancing the magnetic anisotropy as shown in Ref.~\cite{Bryant2013} (see Supplementary Note~1 and Supplementary Fig.~S1).

We account for scattering up to third order in the matrix elements by additionally considering an AFM spin-spin Kondo-exchange coupling $J_{\rm K} \rho_{\rm s}$ between the substrate electrons and the localized spin system \cite{Zhang2013,Ternes2015}, with $\rho_{\rm s}$ the density of states in the substrate around the Fermi energy. When the exchange interaction $J$ between the Co spins is large compared to the Kondo energy $\varepsilon_{\rm K}$, i.e. $|J/\varepsilon_{\rm K}|\gg1$, we find very good agreement between model and experimental data when using a constant exchange coupling of $J_{\rm K} \rho_{\rm s}=0.15$. This value is in accordance with previous results \cite{Delgado2014,Spinelli2014,Oberg2014}. Additional parameter values are listed in Supplementary Table S1.

This model cannot account for the additional spectral weight which arises when the localized spin system enters a state in which it is strongly screened by the substrate electrons. For example, the zero-field Kondo peak of a single Co atom (Fig.~\ref{Figure3}a) is strongly underestimated in the simulations. The observed discrepancy arises most likely from the fact that the model accounts only for correlations induced by scattering events up to third order during the tunnelling process, meaning that processes involving interaction with more than a single substrate electron are not taken into account \cite{Ternes2015}. In a full theoretical treatment of the Kondo interaction, higher order scattering between the localized spin and the substrate electrons produces correlations which strongly influences the spectral weight around the Fermi energy, leading to an increased tunnelling probability and therefore a higher conductance at zero bias.

The difference between measured and simulated spectra provides an easily accessible quantitative measure for the degree of Kondo screening in the system. In Fig.~\ref{Figure3}g we have plotted the mean square error (MSE), defined as ${\rm MSE}=\frac{1}{n}\sum_{n} {\left(G_{\rm exp}-G_{\rm sim}\right)}^2$ (with $n$ the number of data-points in the voltage range and $G_{\rm exp}$ and $G_{\rm sim}$ respectively the normalized measured and simulated conductance), as a function of $J/\varepsilon_{\rm K}$. The MSE, which was determined for spectra at zero field, 4~T and 8~T in the voltage range from $-3$~mV to $3$~mV, scales with the areas coloured in yellow in Figs.~\ref{Figure3}a--f (see inset Fig.~\ref{Figure3}g). We find that for larger field values, the region of largest MSE shifts in the direction of positive $J/\varepsilon_{\rm K}$, as expected based on the phase diagram in Fig.~\ref{Figure1}.

\begin{figure}[htbp]
\begin{centering}
\includegraphics{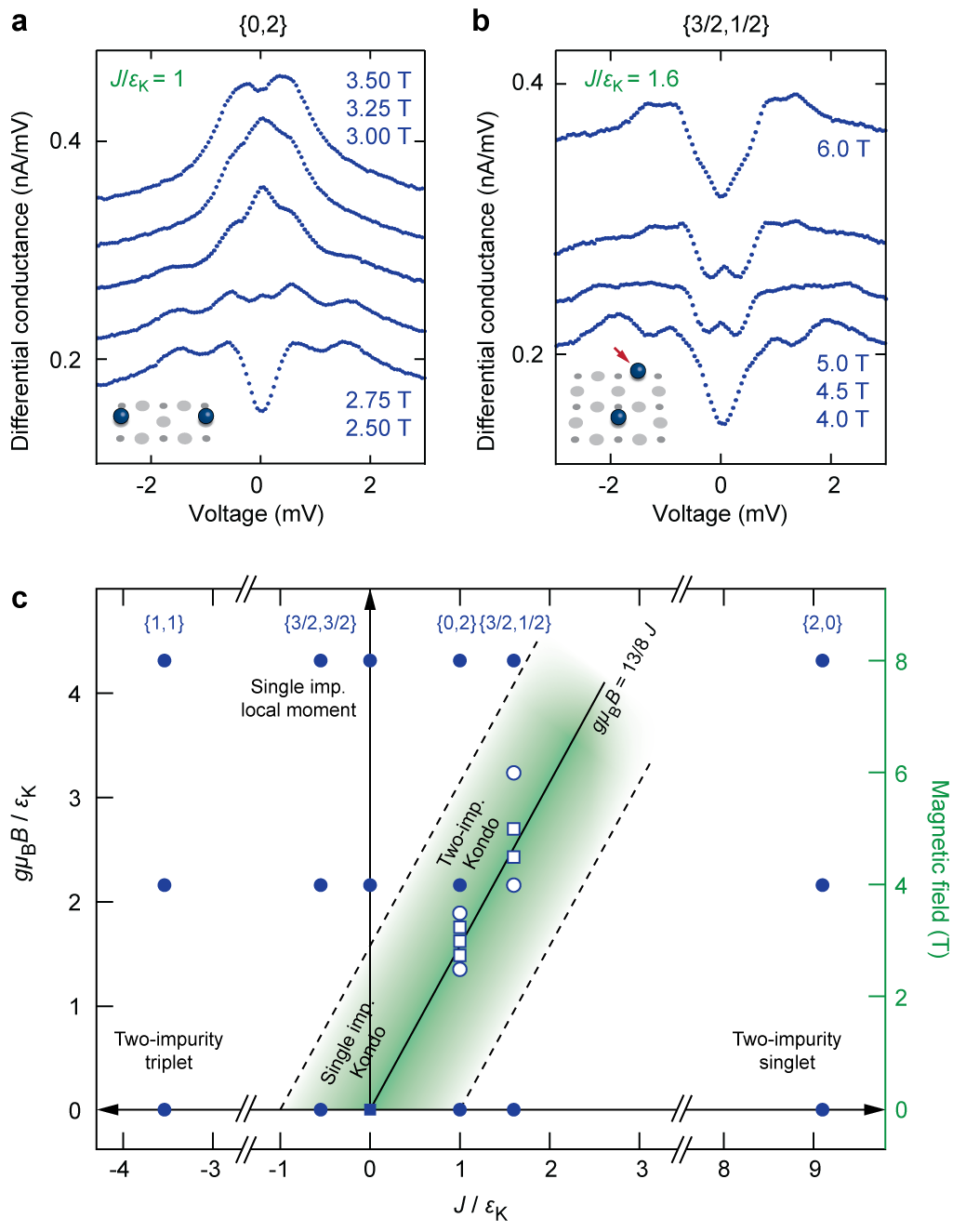}
\caption{\textbf{Two-impurity Kondo screening phase.} (\textbf{a--b}) Differential conductance spectra for the $\left\{0,2\right\}$ and the $\left\{3/2,1/2\right\}$ dimer respectively, around the critical field of each pair. The measurement settings are the same as in Fig.~\ref{Figure3}. (\textbf{c}) Phase diagram of Fig.~\ref{Figure1} showing the locations of measurements presented in Fig.~\ref{Figure3} (blue markers) and panels a and b of this figure (white markers). The data points marked with a square indicate the presence of a Kondo resonance. The solid line marks the position of the ground state degeneracy.}
\label{Figure4}
\end{centering}
\end{figure}

Intriguingly, in the case of AFM coupling, the combined ground state at zero field is the non-magnetic singlet which we can deliberately tune to different phases by applying an external magnetic field. The field effectively counteracts the exchange interaction leading to a crossing of the singlet state with the lowest triplet state at the critical field $B_{\rm c}$. Around this degeneracy point, the system enters a new phase in which the two impurities are together Kondo-screened by the substrate electrons (see Fig.~\ref{Figure1}). Among the dimers studied, there are three with AFM coupling. For the $\left\{2,0\right\}$ dimer the coupling is so strong that the state crossing occurs at a field much larger than 9~T, not accessible in our experimental setup. For the other two cases, $\left\{0,2\right\}$ and $\left\{3/2,1/2\right\}$, the crossings are at approximately 3~T and 5~T, respectively. 

In Figs.~\ref{Figure4}a--b we show a series of ${\rm d}I/{\rm d}V$ spectra measured around these crossing points for the two dimers, in the voltage range from $-3$~mV to $3$~mV. Here, we clearly observe the emergence of Kondo screening concomitant with a zero-voltage peak in the ${\rm d}I/{\rm d}V$ spectrum, which disappears quickly at fields larger or smaller than $B_{\rm c}$. Note that this two-impurity Kondo state differs significantly from that of a single atom at zero field (Fig.~\ref{Figure3}a): it reflects the combined screening of the dimer by the substrate electrons. In addition to the central Kondo peak the spectra show low-energy features near the crossing. These features, that cannot be captured by our current model, may contain additional information about the two-impurity Kondo screening phase. Due to some of these features being close to the central peak, it is difficult to extract the exact Kondo line width. However, in both cases the Kondo peak is clearly wider than the thermal energy of our measurement, which is approximately 30~$\mu$eV.

Fig.~\ref{Figure4}c shows the locations of all presented measurements in the phase diagram of Fig.~\ref{Figure1}. All possible phases are found in our data. The ferromagnetic $\left\{1,1\right\}$ dimer is always in the two-impurity triplet state, while the antiferromagnetic $\left\{2,0\right\}$ dimer is always a two-impurity singlet; for those two instances, no Kondo screening is observed. The weakly FM-coupled $\left\{3/2,3/2\right\}$ dimer is very similar to the single Co: at zero field it behaves like a single Kondo-screened impurity, whereas for large fields the two atoms are separate paramagnetic moments. Lastly, the $\left\{0,2\right\}$ and the $\left\{3/2,1/2\right\}$ both undergo a phase transition from the singlet-phase to being separate moments, through the two-impurity Kondo screening phase. 

In summary, we have been able to experimentally explore the complete phase space of a system of two coupled Kondo atoms in a magnetic field. We showed that for AFM coupled dimers the phase can be continuously tuned from the two-impurity singlet to the single impurity local moment phase solely by an external magnetic field, leading to the formation of the correlated two-impurity Kondo screening phase at the critical crossover field. This transition, which is fundamentally different from the previously investigated zero-field transition between the single impurity Kondo screening and two-impurity singlet phases \cite{Bork2011}, technically  constitutes a quantum phase transition, even though our experiments do not allow us to reveal the associated quantum critical behavior \cite{Jones1989}. Using a third-order perturbative transport model, we have demonstrated that the spectra can be well reproduced except in the vicinity of the ground state crossings, where additional weight around zero bias voltage occurs. These results may form the basis for future work on one and two-dimensional engineered Kondo lattices.

\section{Methods}
\subsection{Experimental setup}
The experiments were performed in a commercial STM system (Unisoku USM 1300S), at low-temperature (330~mK) and in ultra-high vacuum ($<2\times 10^{-10}$~mbar). Magnetic fields up to 8~T were applied perpendicular to the sample surface and the hard uniaxial anisotropy of the Co atoms. The Cu$_2$N/Cu(100) sample was prepared in situ by N$_2$ sputtering on the clean Cu crystal \cite{Leibsle1993}; Cu atoms were evaporated on the precooled Cu$_2$N. The STM tip, made of PtIr, was prepared by indentation in the bare Cu surface. Co dimers were built via vertical atom manipulation \cite{Hirjibehedin2006}. ${\rm d}I/{\rm d}V$ spectra were recorded with a non-polarized tip, using a lock-in technique with excitation voltage amplitude of 70-100~$\mu$V$_{\rm RMS}$ at 928~Hz. The tunnel current for spectroscopy was set between 1.5~nA and 2~nA at -10~mV or -15~mV sample bias for different types of dimer. Variation of the distance between the tip and the atoms up to 150~pm (corresponding to a factor $\sim30$ variation in tunnel current) on the same atom did not lead to any significant modification of the spectral lineshapes. The resolution of the spectral features is limited mostly by the temperature and the lock-in modulation.

\subsection{Simulations}
The simulation of the differential conductance curves was performed using a 
perturbative approach established by Appelbaum, Anderson, and Kondo 
\cite{Kondo1964, Appelbaum1966, Anderson1966, Appelbaum1967} which accounts for 
spin-flip scattering processes up to 3rd order in the matrix elements. The 
transition probability $W_{i\rightarrow f}$ for an electron to tunnel 
between tip and sample or vice versa and simultaneously changing the quantum 
state of the dimer system between the initial ($i$) and final ($f$) state is 
in this model given by:
\begin{equation}
 W_{i\rightarrow f}\propto\left( 
|M_{i\rightarrow
f}|^2+J_{\rm K}\rho_s\sum_{m} \left( \frac{M_{i\rightarrow
m}M_{m\rightarrow
f}M_{f\rightarrow i}}{\varepsilon_i-\varepsilon_m} +\mbox{c. c.}\right) \right)
\delta(\varepsilon_i-\varepsilon_f).
\label{equ:W}
\end{equation}

In this expression, $M_{i\rightarrow j}=\langle 
\psi_j,\sigma_j|(\mathbf{S}\cdot$\boldmath$\sigma$\unboldmath$+u)|\psi_i,
\sigma_i\rangle$ is the scattering matrix element from the combined state vector $|\psi_i,\sigma_i\rangle$ to $|\psi_j,\sigma_j\rangle$, with $\psi$ as the wavevector of the interacting electron and $\sigma$ as the eigenstate of the localized spin system. Energy conservation between initial energy $\varepsilon_i$ and final energy 
$\varepsilon_f$ is obeyed by the delta distribution $\delta(\varepsilon_i-\varepsilon_f)$ in equation~(\ref{equ:W}).

The first term of equation~(\ref{equ:W}) is responsible for the conductance steps 
observed in the spectra \cite{Hirjibehedin2007}, while the second term leads to 
logarithmic peaks at the intermediate energy $\varepsilon_m$ and scales with the 
coupling $J_{\rm K}\rho_s$ to the substrate \cite{Zhang2013}. Note, that this 
model cannot cover strong correlations because it neglects higher order 
effects.

\bibliographystyle{naturemag_noURL}
\bibliography{library}

\subsection{Acknowledgements}
This work was supported by the Dutch Organizations FOM and NWO (VIDI) and by the Kavli Foundation. M.T. acknowledges support by the SFB 767.

\newpage
\section{Supplementary Note 1}
\subsection{Magnetic anisotropy}

Within each type of dimer except for the $\left\{2,0\right\}$, extracted anisotropy parameters were found to be consistent within $\pm5\%$, which is similar to the typical variation found in individual Co atoms. This variation may be attributed to local strain caused by subsurface defects \cite{Otte2008}. However, as shown in Fig.~S1, the $\left\{2,0\right\}$ dimer was found to present much larger variations in the spectra. Interestingly, we could reproduce all the measured curves by varying only one fitting parameter, the anisotropy along the nitrogen direction $\Lambda_{zz}$, and keeping all the other parameters constant. We note that in the conventional notation of $D$ and $E$, two parameters would have to be varied in order to model the observed behaviour. No correlation was found between the $\Lambda_{zz}$ values and either the distance of the dimer from the edge of the Cu$_2$N island, or the size of the supporting island. In Fig.~S1c two examples of $\left\{2,0\right\}$ dimers are shown, that exhibit two extreme values of $\Lambda_{zz}$, but that were built on the same island and at very similar distances from the edges.

The large variation in the anisotropy of the $\{2,0\}$ dimer can be accounted for since the lattice distortion is in a critical regime. It was noted for Fe dimers in this configuration \cite{Bryant2013} that the anisotropy is increased compared to the single atom case, due to an increase in the N-Co-N angle. For Co $\{2,0\}$ dimers this angle approaches $180^{\circ}$: at this critical angle, the orbital levels become degenerate. Since $\Lambda_{zz}$ is inversely proportional to the orbital splitting, close to the critical angle small strain-induced modifications of the crystal field can give rise to major changes in the anisotropy parameters. While the atoms in all other dimers are described as easy-plane systems with the hard axis along the vacancy direction (i.e. $\Lambda_{xx}< \Lambda_{yy}\sim\Lambda_{zz}$), for some instances of the $\left\{2,0\right\}$ dimer $\Lambda_{zz}$ has increased to such an extent that $\Lambda_{zz}-\Lambda_{yy}>\Lambda_{yy}-\Lambda_{xx}$, which classifies as an easy-axis system along the nitrogen direction. When written in terms of conventional phenomenological anisotropy parameters $D$ and $E$, technically dimer IV should be specified as $D=+3.99$~meV, $E=1.26$~meV, and dimer V as $D=-4.23$~meV, $E=1.37$~meV. The smooth variation in the spectroscopy of Fig.~S1a illustrates that it is more natural to use the description in terms of $\Lambda_{\mu\mu}$.

\newpage
\renewcommand\thefigure{\thesection.\arabic{figure}}
\makeatletter 
\renewcommand{\thefigure}{S\@arabic\c@figure}
\makeatother
\section{Supplementary Figure 1}
\setcounter{figure}{0}
\begin{figure}[htbp]
\begin{centering}
\includegraphics{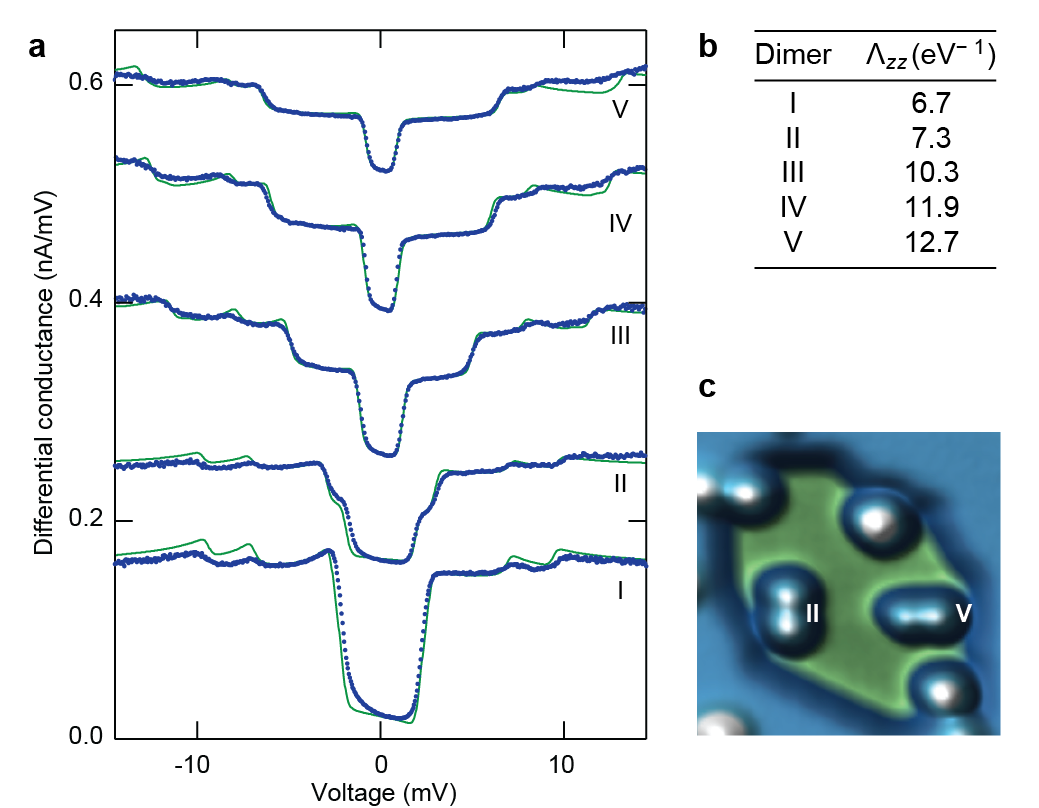}
\caption{\textbf{Anisotropy variation in strained dimers.} (\textbf{a}) Zero-field differential conductance spectra (blue dots) and corresponding simulated curves (green lines) recorded on different instances of a $\left\{2,0\right\}$ dimer, showing variation in the spectroscopic features. (\textbf{b}) Best fit values for $\Lambda_{zz}$ for all dimers shown in \textbf{a}; we found $\Lambda_{xx}=0$, $\Lambda_{yy}=6.2$~eV$^{-1}$, $J=2$~meV and $J_{\rm K}\rho_{\rm s}=0.15$ for all measured curves. (\textbf{c}) Topographic STM image of dimers II and V. These data also published in Ref.~\cite{Bryant2015}.}
\end{centering}
\end{figure}

\newpage
\renewcommand{\figurename}{TABLE}
\setcounter{figure}{0}
\section{Supplementary Table 1}
\begin{figure}[htbp]
\begin{centering}
\includegraphics{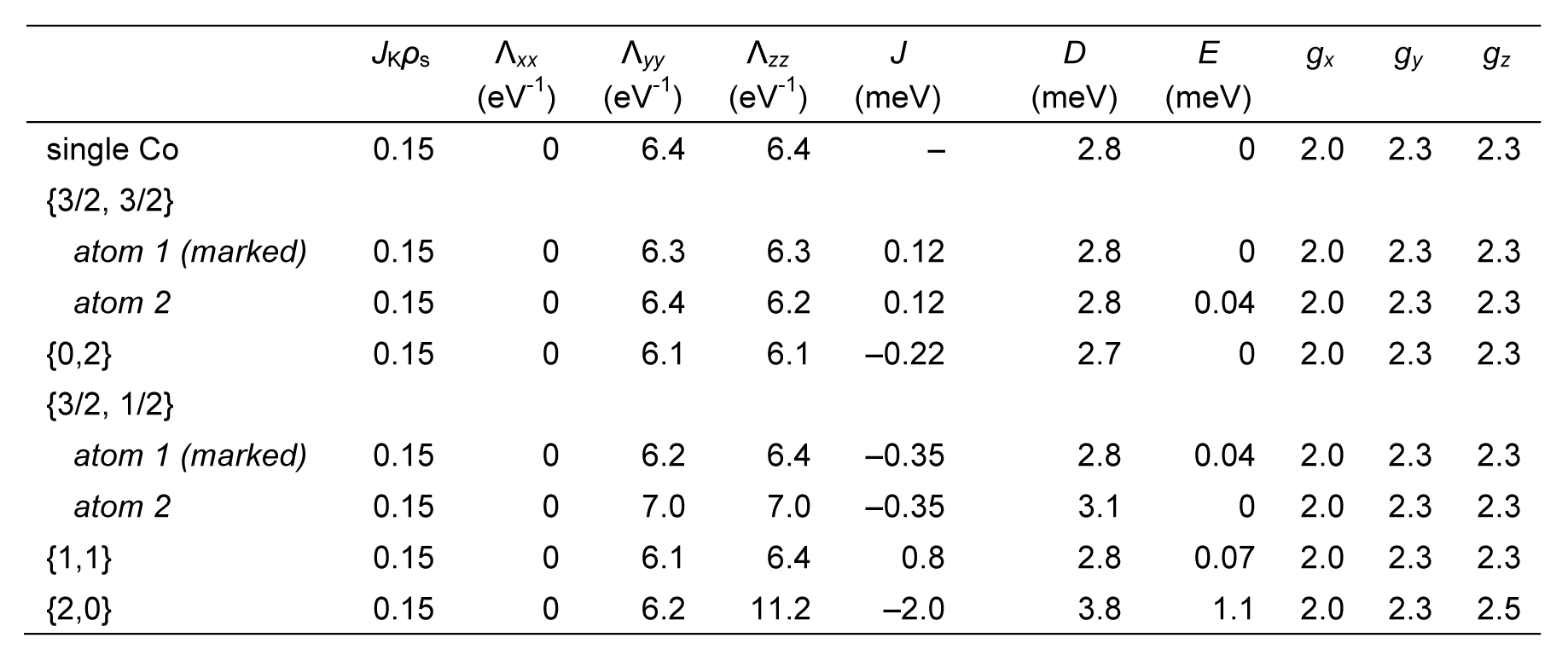}
\caption{\textbf{Fitting parameter values.} Parameter values used for the calculated curves in Fig.~3a--f. Atoms in dimers with inequivalent atom sites (\{3/2,3/2\} and \{3/2, 1/2\}) are marked in Fig.~3 with a red arrow. Parameters $D$, $E$, $g_x$, $g_y$ and $g_z$ are calculated from $\Lambda_{xx}$, $\Lambda_{yy}$ and $\Lambda_{zz}$ and are therefore not independent.}
\end{centering}
\end{figure}

\end{document}